\documentclass{aa}  
\usepackage{graphicx}
\usepackage{txfonts}
\usepackage{natbib}
\bibpunct{(}{)}{;}{a}{}{,} % to follow the A&A style

\usepackage[normalem]{ulem}
\usepackage{color}

\begin{document}

   \title{Rotation of ZZ~Ceti stars as seen by TESS}

   %\subtitle{}

    \author{Zs\'ofia~Bogn\'ar\inst{}\thanks{\email{bognar.zsofia@csfk.org}},
         \'Ad\'am~S\'odor\inst{}}
   
   \institute{
        Konkoly Observatory, HUN-REN Research Centre for Astronomy and Earth Sciences, MTA Centre of Excellence, H-1121 Budapest, Konkoly Thege Mikl\'os \'ut 15-17, Hungary
        }
        
    \titlerunning{ZZ~Ceti rotation}
        \authorrunning{Zs.~Bogn\'ar et al.}
        
    \date{}

% \abstract{}{}{}{}{} 
% 5 {} token are mandatory
 
  \abstract
   {Knowing the rotation rates and masses of white dwarf stars is an important step towards characterising the angular momentum transport mechanism in their progenitors, and coupling the  cores of red giants to their envelopes. However, deriving these rotation rates is not an easy task. One can use the rotational broadening of spectral lines, but there is another way to gather reliable information on the stellar rotation periods of pulsators: through studying the splitting effect of rotation on oscillation frequencies.}
   {We aim to derive stellar rotation periods in the TESS sample for as many white dwarf pulsators as possible.}
   {We rely on light-curve analysis of the TESS observations, and search for closely spaced frequency multiplets that could be rotationally split pulsation modes. We work with triplet frequencies, even if one or two triplet components are only marginally detectable. We also utilise ground-based observations available from the literature in an attempt to confirm the presence of several triplets.}
  {We successfully identified rotationally split multiplets and derived rotation rates for 14 stars. The fastest rotators we identified have rotation periods of 6.6--10.0\,h. The majority of the pulsators rotate with 
periods of 11.9--47.5\,h, while we derived 85.5 and 93.2\,h periods for the slowest rotators. In addition to providing stellar mass estimations, our results confirm previous findings that larger-mass WDs rotate faster than their lower-mass counterparts. We determine the rotation periods of four stars for the first time.}
   {}


   \maketitle
%
%________________________________________________________________

\section{Introduction}

ZZ~Ceti or DAV stars are short-period ($P\sim100-1500$\,s), low-amplitude ($A\sim0.1\%$) white dwarf pulsators with atmospheres dominated by hydrogen. Their pulsation modes are low-spherical-degree ($\ell = 1$ and $2$), and low-to-mid radial-order $g$-modes. They are situated in the 10\,500--13\,000\,K effective-temperature range. The $\kappa-\gamma$ mechanism \citep{1981A&A...102..375D, 1982ApJ...252L..65W} in combination with the convective driving mechanism \citep{1991MNRAS.251..673B, 1999ApJ...511..904G} is responsible for the excitation of the observed pulsations, and they form the most populous group of pulsating white dwarf stars.  

Over time, a large amount of interesting information regarding ZZ~Ceti stars has been revealed. Their pulsation behaviour has long been known to depend strongly on their location within the empirical ZZ~Ceti instability strip. As summarised by \citet{2017ApJS..232...23H}, we find fewer and lower-amplitude pulsation modes in the hot DAVs than in their cooler counterparts. However, the frequency analysis of the cooler DAVs often poses a real challenge: as \citet{2017ApJS..232...23H} described based on Kepler space telescope observations, above about $800\,$s in period, we often find a large number of peaks under a broad envelope instead of one single peak at a given frequency. This phenomenon is reminiscent of stochastically driven oscillations \citep{2015ApJ...809...14B}. Another challenge is to find an explanation for the so-called outburst episodes observed in cool DAV stars close to the red edge of the empirical instability strip. An outburst is an average brightness increase of at least several per cent on a relatively short timescale (of about 1 hour). The star remains in this state for between several hours and one day, before the stellar brightness decreases back to the initial value. The outburst events usually recur after several days or weeks, and sometimes after even longer breaks, such as in the case of HE~0532-5605, as reported by \citet{2023A&A...674A.204B}. We cannot predict the duration of outbursts or the time of their occurrence. Further details on this phenomenon are given in the papers of \citet{2015ApJ...809...14B, 2016ApJ...829...82B, 2017ASPC..509..303B}, and \citet{2015ApJ...810L...5H}. For possible theoretical explanations, including non-linear mode coupling, radiative damping, or phase shifts of the travelling waves reflected from the outer turning point being close to the convection zone, see, for example, the papers of \citet{2017ASPC..509..303B}, \citet{2018ApJ...863...82L}, and \citet{2020ApJ...890...11M}. For reviews of the theoretical and observational aspects of studies of white dwarf pulsators, we also recommend the papers of \citet{2008ARA&A..46..157W}, \citet{2008PASP..120.1043F}, \citet{2010A&ARv..18..471A}, \citet{2019A&ARv..27....7C}, and \citet{2020FrASS...7...47C}.

In the present paper, we study the rotation rates of DAV stars observed by the Transiting Exoplanet Survey Satellite (TESS; \citealt{2015JATIS...1a4003R}). TESS was launched on 18 April 2018, and the main goal of the mission is to find exoplanets at bright nearby stars with the transit method. However, the time sampling of the observations also allows us to follow all kinds of brightness variations of the stars in the observed fields. The two-year primary mission provided not only 30 minute sampling of full-frame images from almost the entire sky (long cadence), but also short-cadence (120 seconds) observations of selected targets. In the case of the short-period white dwarf and subdwarf pulsators, this latter mode provided suitable data sampling for analysis; see for example the first-light papers of the TESS Asteroseismic Science Consortium (TASC) Compact Pulsators Working Group (WG\#8): \citet{2019A&A...632A..42B}, \citet{2019A&A...632A..90C}, and \citet{2020A&A...638A..82B}. Furthermore, in the case of the Extended Mission approved for 2020--2022, new 20 second ultrashort-cadence mode observations became available, which provide an excellent opportunity to study the hot ZZ~Ceti stars with the shortest pulsation periods. Utilising these ultrashort-cadence measurements, several new candidate pulsation modes were detected in some of the southern ecliptic ZZ~Ceti stars \citep{2022MNRAS.511.1574R,2023A&A...674A.204B}.

The present work focuses on the measurement of rotation rates in ZZ Ceti stars. 
In general, we can derive the projected equatorial velocity of a given star using the rotational broadening of spectral lines. However, we face difficulties in applying this method to white dwarf stars. The strong atmospheric pressure broadens their line profiles, masking the relatively small contribution of stellar rotation to the overall broadening. However, we can find examples in the literature of attempts to derive the $v\,\mathrm{sin}\,i$ values for some stars; see \citet{1998A&A...338..612K} and \citet{2005A&A...444..565B}. There is another way to determine the rotational period of a white dwarf, which involves the use of inhomogeneities on the rotating stellar surface, such as spots caused by the star's magnetic field or variations in chemical composition (see e.g. \citealt{2013ApJ...773...47B} and \citealt{2000PASP..112..873W}).
Finally, there is a third way to gather reliable information on the stellar rotation periods for pulsators: through studying the splitting effect of rotation on oscillation frequencies. The separation of these frequency components highly depends on the rotation rate in the stellar layer probed by the given pulsation mode, which offers a unique opportunity to reveal their internal rotation. The white dwarf rotation rates as a function of mass may shed light on the unknown angular momentum transport mechanism in their progenitors, coupling red-giant cores to their envelopes \citep{2017ApJS..232...23H}. For more details, we also recommend the papers of \citet{2003astro.ph..1539K}, \citet{2015ASPC..493...65K}, and \citet{2019A&ARv..27....7C}. As we need long-time-base and precise observations to resolve the often closely spaced and low-amplitude frequency components, space-based data provide the best opportunity to obtain such measurements.

\section{Analyses of the TESS data sets}
\label{sect:analyses}

Even though TESS observed hundreds of potentially pulsating DA white dwarf stars, here we do not intend to search for new DAV pulsators. We investigated the already known ZZ~Ceti stars instead. For this purpose, we relied on two databases of known DAVs: \citet[][Table~4]{2016IBVS.6184....1B} and \citet[][Table~1]{2022MNRAS.511.1574R}.

TESS observed 91 of the 180 stars listed in Table~4 of \citet{2016IBVS.6184....1B}, while the work of \citet{2022MNRAS.511.1574R} is based on TESS observations, and therefore TESS light curves are available for all 74 new DAV stars presented in this latter work. Our final sample includes 165 DAVs with available short-cadence TESS light curves. We note that we investigated the TESS data sets up to and including sector~55 (cycle~4).

Our main goal is not to perform a detailed frequency analysis of the TESS data sets of every star in our sample, but to search for possible rotationally split frequency components (preferably triplets), and to use them for measuring the rotation rates of these stars.

We began by downloading all the light curves of our sample from the Mikulski Archive for Space Telescopes (MAST), and extracting the PDCSAP fluxes provided by the presearch data-conditioning pipeline \citep{2016SPIE.9913E..3EJ}. This pipeline corrects the flux of each target to account for the contamination of nearby stars caused by the frequent crowding. Next, we corrected the light curves for long-term systematic trends, and removed outlier measurements. We divided the time strings into segments containing gaps no longer than 0.5~d. We then separately fitted and subtracted cubic splines from each segment. We used one knot point for every 200 (short-cadence mode) or 1000 points (ultrashort-cadence mode) to define the splines. Finally, we removed outliers that showed a deviation from the mean brightness of greater than 4 sigma. Considering that the removed trends correspond to significantly longer timescales than the DAV pulsation periods, these corrections do not affect the frequency domain of the white dwarf pulsations. We note that when observations were available from multiple sectors, we analysed the complete, merged light curves together.

Afterwards, we analysed the corrected data sets with the command-line light-curve fitting program \textsc{LCfit} developed by \'A. S\'odor \citep{2012KOTN...15....1S}. Utilising an implementation of the Levenberg-Marquardt least-squares fitting algorithm, \textsc{LCfit} is capable of linear (amplitudes and phases) and non-linear (amplitudes, phases, and frequencies) least-squares fittings. The program handles unequally spaced measurements, and data sets with intermittent gaps.

Finally, we inspected the obtained pulsation data. We searched for signs of regularities in the frequency spacings, preferably looking for equidistant triplets, but we also checked closely spaced frequency doublets. We accepted a frequency component as significant if its amplitude exceeded five times the local noise level in the periodogram (S/N\,>\,5). We also checked the Lomb-Scargle periodograms produced by the \textsc{LCfit} program by eye in the case of stars showing probable frequency splittings. We obtained a total of 14 stars showing frequency triplets or doublets for further investigation.

\section{Doublets and triplets}
\label{sect:multiplets}

\begin{figure*}
\centering
\includegraphics[width=1.0\textwidth]{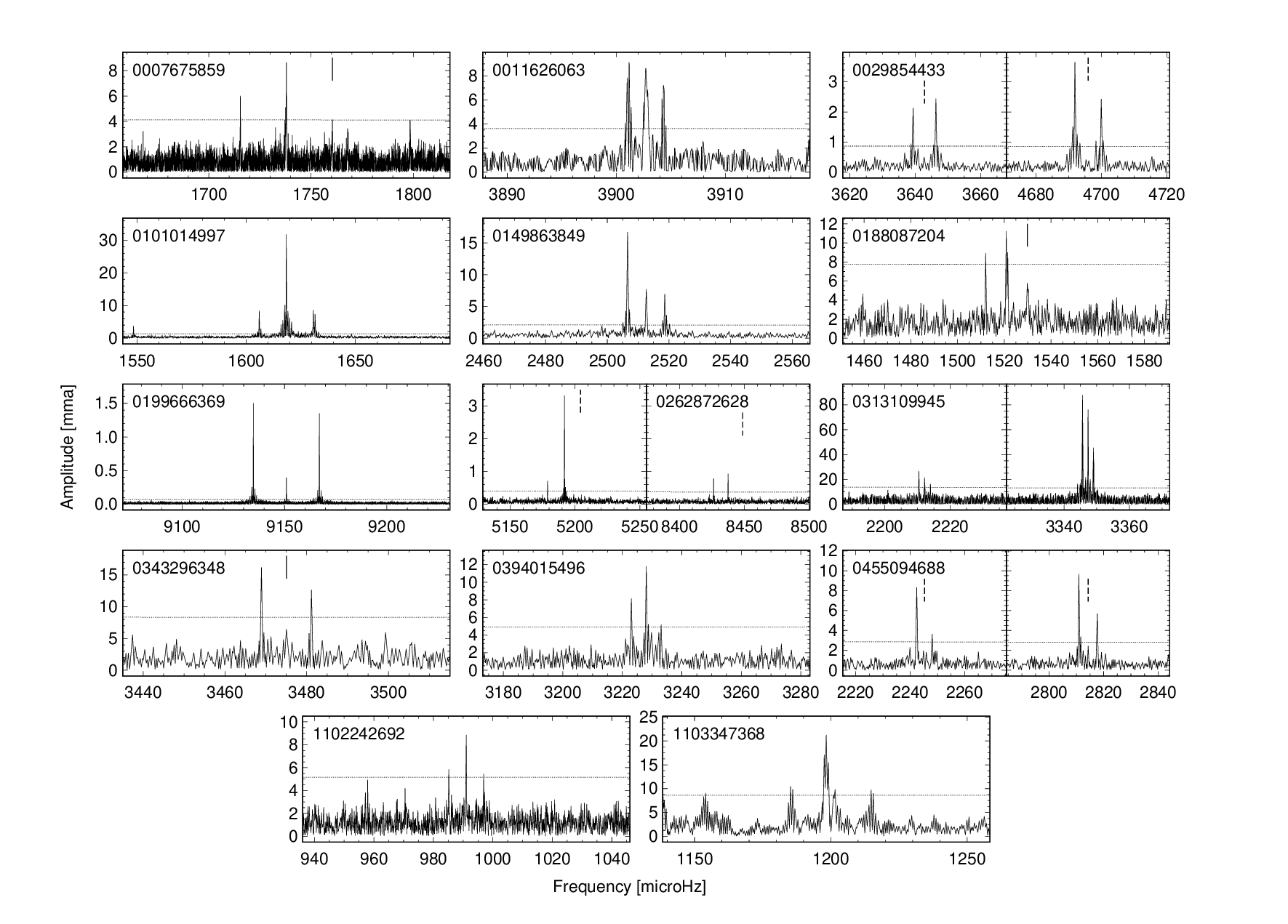}
\caption{Fourier transforms of the 14 stars showing doublet or triplet frequencies. Horizontal dotted lines denote the five signal-to-noise-ratio significance levels. Detected triplet components slightly below this level are marked with continuous markers, and the locations of undetected triplet components known from the literature are marked with dashed markers.}
\label{fig:multiplets}
\end{figure*}

\begin{table*}
\centering
\caption{List of ZZ~Ceti stars with triplet or doublet frequencies found by the frequency analysis of the TESS light curves. \textit{TIC ID} refers to the TESS Input Catalog identifier of the object. The frequencies and the corresponding frequency separations of the components are also denoted.}
\label{tabl:multiplets1}
\begin{tabular}{lcrl}
\hline
\hline
\multicolumn{1}{c}{TIC ID} & \multicolumn{1}{c}{Frequencies of multiplets} & \multicolumn{1}{c}{$\delta\nu$} & \multicolumn{1}{c}{Comments}\\
 & \multicolumn{1}{c}{[$\mu$Hz]} & \multicolumn{1}{c}{[$\mu$Hz]} & \\
\hline
0007675859 & 1715.42 & & \\
           & 1737.91 & 22.19 & \\
           & 1760.42 & 22.51 & doublet, with a low-amplitude third component\\
0011626063 & 3901.16 & & \\ 
           & 3902.75 & 1.59 & \\
           & 3904.32 & 1.57 & triplet\\
0029854433 & 3639.34 & & \\
           & 3646.31 & 6.97 & \\
           & 4691.92 & & \\
           & 4699.94 & 8.02 & two doublets, side components of triplets\\
0101014997 & 1605.95 & & \\
           & 1618.37 & 12.42 & \\
           & 1630.78 & 12.41 & triplet\\
0149863849 & 2506.60 & & \\
           & 2512.65 & 6.05 & \\
           & 2518.64 & 5.99 & triplet\\
0188087204 & 1512.05 & & \\
           & 1520.85 & 8.80 & \\
           & 1529.86 & 9.01 & doublet, with low-amplitude third component\\
0199666369 & 9134.73 & & \\
           & 9150.86 & 16.13 & \\
           & 9167.00 & 16.14 & triplet\\
0262872628 & 5178.86 & & \\
           & 5191.85 & 12.99 & \\
           & 8426.32 & & \\
           & 8437.44 & 11.12 & two doublets, adjacent components of triplets\\
0313109945 & 2210.42 & & \\
           & 2212.11 & 1.69 & \\
           & 2213.92 & 1.81 & \\
           & 3345.66 & & \\
           & 3347.36 & 1.70 & \\
           & 3349.05 & 1.69 & two triplets\\
0343296348 & 3468.97 & & \\
           & 3475.02 & 6.05 & \\
           & 3481.15 & 6.13 & doublet, with a low-amplitude middle component\\
0394015496 & 3223.02 & & \\
           & 3228.04 & 5.02 & \\
           & 3233.02 & 4.98 & triplet\\
0455094688 & 2242.39 & & \\
           & 2248.07 & 5.68 & \\
           & 2810.92 & & \\
           & 2817.63 & 6.71 & two doublets, side components of triplets\\
1102242692 & 985.19 & & \\
           & 991.06 & 5.87 & \\
           & 996.91 & 5.85 & doublet, with low-amplitude third component\\
1103347368 & 1185.24 & & \\
           & 1198.33 & 13.09 & \\
           & 1214.77 & 16.44 & triplet\\
\hline
\end{tabular}
\end{table*}

Figure~\ref{fig:multiplets} shows the Fourier transforms calculated by \textsc{Period04} \citep{2005CoAst.146...53L} of the 14 stars with identified frequency multiplets, while Table~\ref{tabl:multiplets1} summarises the frequencies detected in these DAVs. The frequency separations of the components are also listed, as we used these to derive the rotation rates.

A plausible explanation for the observed triplet structures is that these are rotationally split frequency components of $l=1$ horizontal-degree modes. Knowing the frequency differences of the triplet components ($\delta \nu$), we can estimate the rotation period of the pulsator.

In the case of slow rotation, the frequency differences of the various azimuthal-order modes of $m=-1,0,1$ rotationally split components can be calculated (to first order) using the following relation: 

\begin{equation}
\label{eq:rot}
\delta \nu_{k,\ell,m} = m (1-C_{k,\ell}) \Omega, 
\end{equation}
\noindent where $k$, $l$, and $m$ are the radial order, horizontal degree, and azimuthal order of the non-radial pulsation mode, respectively, the coefficient \mbox{$C_{k,\ell} \approx 1/\ell(\ell+1)$} for high-overtone ($k\gg\ell$) $g$-modes, and $\Omega$ is the (uniform) rotational frequency.

However, as described for example in \citet{2017ApJS..232...23H}, we cannot assume that all the observed doublet and triplet frequencies are high overtone modes. For the sake of comparison with the rotation rate results presented by \citet{2017ApJS..232...23H}, we uniformly accepted a value of $C_{k,\ell} = 0.47$ for estimating the rotation periods for each object. Table~\ref{tabl:multiplets2} summarises the results for our  sample of 14 stars. For the calculations of rotation rates, we used the average value of the frequency separations ($\delta\nu_{\mathrm{average}}$).

\begin{table}
\centering
\caption{Rotation rates as calculated using Eq.~1 and stellar masses of the 14 selected stars. \textit{TIC ID} refers to the TESS Input Catalog identifier of the object.}
\label{tabl:multiplets2}
\begin{tabular}{lrrr}
\hline
\hline
\multicolumn{1}{c}{TIC ID} & \multicolumn{1}{c}{$\delta\nu_{\mathrm{average}}$} & \multicolumn{1}{c}{P$_{rot}$} & \multicolumn{1}{c}{$M_*$} \\
 & \multicolumn{1}{c}{[$\mu$Hz]} & \multicolumn{1}{c}{[h]} & \multicolumn{1}{c}{[$M_{\odot}$]} \\
\hline
0007675859 & 22.35 & 6.6 & 0.91 \\
0011626063 & 1.58 & 93.2 & 0.65 \\
0029854433 & 3.7475 & 39.3 & 0.62 \\
0101014997 & 12.415 & 11.9 & 0.64 \\
0149863849 & 6.02 & 24.5 & 0.66 \\
0188087204 & 8.905 & 16.5 & 0.41 \\
0199666369 & 16.135 & 9.1 & 0.83 \\
0262872628 & 12.055 & 12.2 & 0.69 \\
0313109945 & 1.7225 & 85.5 & 0.38 \\
0343296348 & 6.09 & 24.2 & 0.59 \\
0394015496 & 5.00 & 29.4 & 0.61 \\
0455094688 & 3.098 & 47.5 & 0.72\\
1102242692 & 5.86 & 25.1 & 0.37 \\
1103347368 & 14.765 & 10.0 & no data\\
\hline
\end{tabular}
\end{table}

As one of our goals with the presented investigations is to check the mean rotation periods for the DAVs at different masses, we also had to derive the mass values of the targets. In the case of the \citet{2022MNRAS.511.1574R} database, the masses for the stars from different sources are listed in their Table~1. However, for the stars selected from the database of \citet{2016IBVS.6184....1B}, only the surface gravity (log\,$g$) values are available, which have to be converted to masses. For this purpose, we used the log\,$g$--$M_*$ relations  based on the evolutionary sequences published by \citet{2020ApJ...901...93B} as presented on the `synthetic colors and evolutionary sequences of hydrogen- and helium-atmosphere white dwarfs' webpage\footnote{\url{http://www.astro.umontreal.ca/~bergeron/CoolingModels}}.

In the followings, we comment on the individual stars we identified as rotators.

%\subsection{Results for individual targets}

\subsection{TIC~0007675859}

We identify a wide doublet, which is possibly a triplet with a third component somewhat below the level of S/N=5. The pulsations of the star were discovered by \citet{2022MNRAS.511.1574R}. We note that this star is one of the fastest rotators found in our investigations. Based on the frequency separations of the presumed triplet, we determine a $6.6$\,h rotation period for the star. We note that Romero et al. (2022; Table~8) present a different frequency triplet. We tried to find this triplet by analysing our data. We lowered the significance level of our frequency search to\,S/N=4.5, which allowed us to identify a doublet with the frequencies of $2830.5$ and $2807.9\,\mu$Hz. The frequencies of the adjacent triplet components in \citet{2022MNRAS.511.1574R} are $2830.9$ and $2808.3\,\mu$Hz, and the corresponding rotation period calculated by these authors is $5.2$\,h ($C_{k,\ell}=0.487$). Assuming $C_{k,\ell}=0.47$, this value would be $5.3$\,h.  \citet{2022MNRAS.511.1574R} may have found a different triplet for the star because of the different data set used: these authors analysed the 120\,s cadence data from Sectors 25--26, while we present results from our analysis of the 20\,s cadence
observations from Sectors 40 and 52. We therefore also analysed the 120\,s cadence data from Sectors 25--26  independently, and find the doublet mentioned above as the first and second highest signals of this data set. We find the third component reported by \citet{2022MNRAS.511.1574R} as a low-amplitude peak. The difference between the frequency content of the different Sectors suggests temporal amplitude variations. This could also be the case for TIC~21187072, where \citet{2022MNRAS.511.1574R} reported a triplet that we do not identify in our data set. However, from our analysis of the 120\,s cadence Sector 25--26 data investigated by \citet{2022MNRAS.511.1574R}, we indeed find a doublet with the frequencies of $928.6$ and $930.8\,\mu$Hz. The third frequency component can also be seen with low amplitude. We present the results of the frequency analysis of the 20\,s cadence observations from Sectors 40, 41, and 52  in the present work, and only find a significant peak at $928.5\,\mu$Hz. TIC~21187072 was not included in our sample because here we only present the results of independent frequency analyses based on the 20\,s cadence observations.

\subsection{TIC~0011626063}

This star is also known as GD~385. There is an unambiguous triplet in the periodogram. \citet{1984ApJ...278..754K} detected two frequencies at around $3900\,\mu$Hz, but did not find a third component. However, these authors highlighted the possibility that the multiplet is the product of rotational frequency splitting. \citet{1984ApJ...278..754K} estimated the value of $\ell$ of these frequencies to be about 2,  and determined an 89\,h rotation period for GD~385. We recalculated the rotation period of the star, assuming that the 3901.22 and 3904.27\,$\mu$Hz frequencies are side components of an $\ell=1$ triplet, and $C_{k,\ell}=0.47$. We obtain $96.7$\,h for the rotation period of GD~385. This value is close to the period determined from the TESS observations presented in the present work ($93.2$\,h), and is also not far from the period calculated by \citet{1984ApJ...278..754K}.

\subsection{TIC~0029854433}

This star is the well-known pulsator Ross~548. We detect two doublets in the TESS data set, but considering previous literature data based on ground-based observations, we know that these are actually side components of two triplets; see for example \citet{2015ApJ...815...56G}. According to the measurements performed by these latter authors for the two triplets, the average rotation period of the star is $39.5$\,h, assuming that $C_{k,\ell}=0.47$. Although \citet{2015ApJ...815...56G} do not present rotation calculations in their publication, the $39.5$\,h period is close to the $39.3$\,h value obtained by our analysis, and is also not far from the period ($37.8$\,h) derived by \citet{2017ApJS..232...23H}.

\subsection{TIC~0101014997}

This star is another well-known variable, BPM~31594. It shows an unambiguous triplet, the structure of which was discovered with ground-based observations by \citet{1992MNRAS.258..415O}, before confirmation with TESS observations; see \citet{2020A&A...638A..82B, 2023A&A...674A.204B}. \citet{1992MNRAS.258..415O} published a $12.7\,\mu$Hz frequency separation for the triplet, and assuming that $m (1-C_{k,\ell}$) is of order unity, these authors derived a $\sim 0.9$\,d ($21.6$\,h) rotation period. Recalculating this with $C_{k,\ell}=0.47$, we obtain a rotation period of $11.6$\,h, which is close to the value we present in this work based on TESS ultrashort-cadence data ($11.9$\,h).

\subsection{TIC~0149863849}

In this star an obvious triplet emerges from the TESS periodogram. The pulsations of the star were discovered by \citet{2022MNRAS.511.1574R}, who also found a triplet structure, which they present in their Fig.~6. However, these authors do not include the corresponding frequency, $C_{k,\ell}$, and mean rotation period values in their Table~8, which is supposed to list all objects with detected rotational splittings. In the present work, we determine a rotation period of $24.5$\,h.

\subsection{TIC~0188087204}

We identify a doublet in the frequency spectrum, which appears to be part of a triplet, with a third frequency component of  low amplitude. The pulsation of the star was also discovered by \citet{2022MNRAS.511.1574R}. Similarly to TIC~0149863849, the authors presented a possible triplet (see their Fig.~6), but the star and the corresponding frequency values are also not included in their Table~8. We obtain a $16.5$\,h rotational period from the TESS measurements.

\subsection{TIC~0199666369}

This star is another well-known pulsator, also known as G~226-29, and is the brightest known DAV star. Previous ground-based observations only showed three frequency components; see for example the results presented by \citet{1995ApJ...447..874K}. Our analysis of the TESS data also confirms this result, showing only the same triplet. \citet{1995ApJ...447..874K} found that the average rotation frequency splitting is $16.15\,\mu$Hz. These authors used $C_{k,\ell}=0.48$ to calculate the rotation period of the star. Their result is $8.9$\,h, while assuming $C_{k,\ell}=0.47$ we obtain $9.1$\,h. We also calculate a rotation period of $9.1$\,h based on the present analysis of the TESS observations.

\subsection{TIC~0262872628}

This pulsator is also known by the name L~19-2. Ground-based observations revealed triplet structures in the light curve of the star; see \citet{2005ApJ...635.1239Y}. Considering the TESS data, we detect only the adjacent components of these triplets as two doublets \citep{2023A&A...674A.204B}. \citet{2005ApJ...635.1239Y} do not present rotational period calculations, only frequency separations. However, using the average of these values and  assuming $C_{k,\ell}=0.47$, we calculate a rotational period of $12.5$\,h, which is close to the $12.2$\,h presented in this work, and also to the $13.0$\,h calculated by \citet{2017ApJS..232...23H}.

\subsection{TIC~0313109945}
\label{asimbadosmagyarazkodas}

We detect two triplets in the TESS light curve. Calculating the average of their frequency separations, and using $C_{k,\ell}=0.47$, we obtain a rotation period of $85.5$\,h. The pulsation of the star was discovered by \citet{2022MNRAS.511.1574R}. We note that according to SIMBAD, TIC~0313109945 is a blend of GD~491 with a G magnitude of 15.6 and a high-proper-motion star with 18.7 G magnitude. The separation is 3 arcsec between the two. The TIC CROWDSAP value of 0.08 therefore corresponds to the fainter background star, and it should be 0.92 relative to our target, because CROWDSAP expresses the flux of the target relative to the total flux.

\subsection{TIC~0343296348}

The TESS data reveal a triplet structure in the periodogram for this star. Its pulsations were discovered by \citet{2022MNRAS.511.1574R}. These authors also found the same triplet, and determined a rotation period of $24.5$\,h. With our calculations, the stellar rotational period is $24.2$\,h, which is fairly close to the \citet{2022MNRAS.511.1574R} solution.

\subsection{TIC~0394015496}

A pronounced triplet appears in the TESS periodogram. The pulsation of the star was discovered by \citet{2022MNRAS.511.1574R}. Similarly to the above case of TIC~0343296348, the calculated rotational periods for this star are almost the same: \citet{2022MNRAS.511.1574R} present a $29.8$\,h period, while we derived $29.4$\,h in this work.

\subsection{TIC~0455094688}

This star is also known as HS~0507+0434B. We see two doublets in the TESS periodogram, but former ground-based observations revealed that these are actually the side components of two triplets \citep{2013MNRAS.429.1585F}. These latter authors detected six triplets with an average frequency spacing of $3.59\,\mu$Hz. The corresponding rotational period presented is $38.7$\,h, assuming that $C_{k,\ell}=0.5$. If $C_{k,\ell}=0.47$ is used, we obtain $41.0$\,h, which is almost the same value as that determined by \citet{2017ApJS..232...23H} ($40.9$\,h).  Based on the TESS data used in this work, we get $47.5$\,h with $C_{k,\ell}=0.47$, which suggests a slightly slower rotation.

\subsection{TIC~1102242692}

We find a triplet structure in the TESS periodogram, although the side components are close to the significance limit. We calculated a rotation period of $25.1$\,h.

\subsection{TIC~1103347368}

We see at least one triplet in the TESS periodogram. The side components of the other possible triplet are below the significance level. We calculate a rotation period of $10.0$\,h based on the average of the frequency spacings in the TESS data.

\vspace{5mm} % Should be one blank line but couldn't figure out how to do that.

These objects are listed with their journal of observations in Table~\ref{tabl:journal}. Ultrashort-cadence-mode observations are available for 12 of the 14 analysed stars. We only used short-cadence observations for our analysis in the remaining two targets, as listed in Table~\ref{tabl:journal}.

We note that \citet{2023MNRAS.526.2846U} published their frequency analysis and modelling result on the TESS light curve of another ZZ~Ceti star, G~29-38 (TIC~422526868). These authors identified rotational frequency triplets, and also a quintuplet in the periodogram, from which the estimated rotation period of G~29-38 is about $1.35 \pm 0.1$\,d. We also detected numerous pulsation peaks in these data, but we were not able to establish rotational multiplets among them, and therefore this star does not appear in our results.

%\begin{table*}
\begin{sidewaystable*}
\centering
\caption{Journal of observations of the 14 DAV pulsators showing doublet or triplet structures in their TESS periodograms. \textit{TIC ID} and \textit{Gaia DR3 ID} refer to the TESS Input Catalog and Gaia Data Release 3 identifiers of the object, respectively, \textit{N} is the number of data points after cleaning the light curve, $\delta T$ is the total length of the data sets including gaps, and \textit{Sect.} is the serial number of the sector(s) in which the star was observed. \textit{US} and \textit{S} denote that we used the ultrashort- or short-cadence-mode observations, respectively. The start time in BJD is the time of the first data point in the data set. The \textit{CROWDSAP} keyword represents the ratio of the target flux to the total flux in the TESS aperture. We note that here we list the CROWDSAP value from the first TESS run, even when data from other sectors are available. (*): we corrected the CROWDSAP value from 0.08, which corresponds to a background star, as detailed in Sect.~\ref{asimbadosmagyarazkodas}.}
\label{tabl:journal}
\begin{tabular}{lcccrrcrc}
\hline
\hline
Object        & TIC ID     & Gaia DR3 ID       &Start time & \multicolumn{1}{c}{\textit{N}} & \multicolumn{1}{c}{$\delta T$} & \textit{G} mag & Sect. (Cadence) & \multicolumn{1}{c}{CROWDSAP} \\
& & & (BJD-2\,457\,000) & & \multicolumn{1}{c}{(d)} & & & \\
\hline
--            & 0007675859 & 2114811453822316160 & 2390.655 &  388881 & 405.5 & 16.3 & 40,52--54 (US)      & 0.29 \\
GD 385        & 0011626063 & 2026653131220381824 & 2419.993 &  194107 & 376.1 & 15.1 & 41,54 (US)          & 0.07 \\
Ross 548      & 0029854433 & 2457759374023232768 & 2115.890 &   91784 &  25.7 & 14.2 & 30 (US)             & 0.98 \\
BPM 31594     & 0101014997 & 4835794942327992704 & 2115.887 &  191458 &  54.1 & 15.1 & 30--31 (US)         & 0.90 \\
--            & 0149863849 & 5961193055261256320 & 2361.776 &  106133 &  27.9 & 13.6 & 39 (US)             & 0.15 \\
--            & 0188087204 & 5470271185153118208 & 2282.141 &   14913 &  23.9 & 16.8 & 36 (S)              & 0.38 \\
G 226-29      & 0199666369 & 1431783457574556672 & 2390.654 & 1023148 & 433.6 & 12.3 & 40,41,48--55 (US)   & 0.98 \\
L 19-2        & 0262872628 & 5772718006135360128 & 2333.857 &  218835 &  55.9 & 13.4 & 38,39 (US)          & 0.78 \\
GD~491        & 0313109945 & 1712016196599965312 & 2390.651 &  506832 & 378.3 & 15.6 & 40,41,47,48,53 (US) & 0.92(*) \\
--            & 0343296348 & 5802990619265616640 & 1629.139 &   50104 & 760.6 & 15.9 & 39 (US)             & 0.26 \\
--            & 0394015496 & 6410501923531836928 & 1325.300 &   46191 & 759.4 & 15.8 & 28 (US)             & 0.39 \\
HS 0507+0434B & 0455094688 & 3238868171156736768 & 1437.990 &   16629 &  25.7 & 15.4 & 5 (S)               & 0.44 \\
WD 1526+558   & 1102242692 & 1601328670269782400 & 2640.430 &  213467 &  77.1 & 17.1 & 49--51 (US)         & 0.93 \\
WD 1521-003   & 1103347368 & 4415701304886958848 & 2698.355 &   39187 &  19.2 & 15.8 & 51 (US)             & 0.12 \\
\hline
\end{tabular}
%\end{table*}
\end{sidewaystable*}

\begin{figure}
\centering
\includegraphics[width=0.49\textwidth]{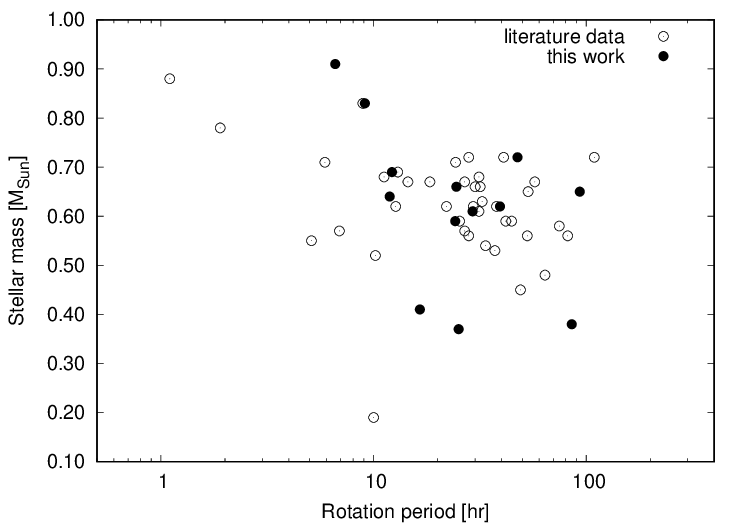}
\caption{Our rotation period results (full dots) compared to literature data (empty circles) versus stellar mass. Literature data were collected from \citet{2017ApJS..232...23H}.}

%Stellar masses are all taken from the literature. Literature data were collected mainly from \citet{2017ApJS..232...23H}, while data on three additional stars are from \citet{2017ApJ...851...24B}, \citet{2021ApJ...922..220L}, and \citet{2023MNRAS.526.2846U}.}
\label{fig:rotmass}
\end{figure}

%%%%%%%%%%%%%%%%%%%%%%%%%%%%%%%%%%%%%%%%%%%%%%%%%%%%%%%%%%%%%%%%%%%%%%%%%%%%%%%%%%%%%%%%%%%%%%%%%%%%%%%%%%%

\section{Summary and conclusions}
\label{sect:disc}

We cannot discuss white dwarf rotation rates in the space age without mentioning the work of J.J.~Hermes and his coworkers published in 2017 \citep{2017ApJS..232...23H}. 
These authors collected results based on ground-based observations and complemented these with their sample based on Kepler and K2 measurements.
One of their findings is that WDs in the 0.51--0.73 solar mass range have a mean rotation period of 35 hours, with a standard deviation of 28 hours.  
They also presented rotation rates as a function of mass in their paper. Another interesting finding by these authors is that the more massive WDs appear to rotate faster; however, further observations are needed to confirm this finding.

\cite{2017ApJS..232...23H} calculated all rotation periods via rotationally split pulsation frequencies detected in ground- and space-based observations. 
We note that all white dwarfs presented in their paper appear to be isolated stars, and so these rotation periods should be representative of the endpoints of single-star evolution.

These results inspired us to perform a similar investigation in the TESS data. Using doublets, we must be cautious when interpreting the rotation rates, as from a doublet frequency we cannot decipher whether or not these are side components or adjacent members of a triplet, or even components of a possible quintuplet or higher-order multiplet. Therefore, we decided to work with triplet frequencies only. However, a pulsation component that is a member of a triplet can provide a strict frequency constraint, and so we somewhat relaxed the amplitude S/N requirement for accepting the detection of some triplet members. Ground-based observations also helped us to confirm whether or not there is a triplet in a given frequency domain.

Another consideration of our work was that we showed a preference for ultrashort-cadence time series, if available for the stars of our sample, as it is expected that frequencies of ZZ~Ceti stars emerge above the Nyquist limit of the 120 second sampling data sets.

\begin{table*}
\centering
\caption{Comparison of the rotation periods obtained as part of the present study with the results published by \citet{2017ApJS..232...23H} (their Table 4), \citet{2022MNRAS.511.1574R} (their Table 8), and \citet{1992MNRAS.258..415O}. \textit{TIC ID} refers to the TESS Input Catalog identifier of the object. If another common identifier is available, it is also given in parentheses. \textit{P$_{rot}$ literature} refers to a measurement from a previous work.}
\label{tabl:multiplets3}
\begin{tabular}{lrrr}
\hline
\hline
\multicolumn{1}{c}{TIC ID (other ID)} & \multicolumn{1}{c}{P$_{rot}$ [h]} & \multicolumn{1}{c}{Reference} & \multicolumn{1}{c}{P$_{rot}$ [h]} \\
 & \multicolumn{1}{c}{literature} & & \multicolumn{1}{c}{this work} \\
\hline
0007675859                 &  5.2\ \ \ \ \ & \citet{2022MNRAS.511.1574R} &  6.6\ \ \ \ \\
0029854433 (Ross 548)      & 37.8\ \ \ \ \ & \citet{2017ApJS..232...23H} & 39.3\ \ \ \ \\
0101014997 (BPM 31594)     & 21.6\ \ \ \ \ & \citet{1992MNRAS.258..415O} & 11.9\ \ \ \ \\
0149863849                 & no data\ \ \ \ \ & \citet{2022MNRAS.511.1574R} & 24.5\ \ \ \ \\
0188087204                 & no data\ \ \ \ \ & \citet{2022MNRAS.511.1574R} & 16.5\ \ \ \ \\
0199666369 (G 226-29)      &  8.9\ \ \ \ \ & \citet{2017ApJS..232...23H} &  9.1\ \ \ \ \\
0262872628 (L 19-2)        & 13.0\ \ \ \ \ & \citet{2017ApJS..232...23H} & 12.2\ \ \ \ \\
0343296348                 & 24.5\ \ \ \ \ & \citet{2022MNRAS.511.1574R} & 24.2\ \ \ \ \\
0394015496                 & 29.8\ \ \ \ \ & \citet{2022MNRAS.511.1574R} & 29.4\ \ \ \ \\
0455094688 (HS 0507+0434B) & 40.9\ \ \ \ \ & \citet{2017ApJS..232...23H} & 47.5\ \ \ \ \\
\hline
\end{tabular}
\end{table*}

\begin{table}
\centering
\caption{Stars not listed either in \citet{2017ApJS..232...23H} (their Table 4) or \citet{2022MNRAS.511.1574R} (their Table 8). \textit{TIC ID} refers to the TESS Input Catalog identifier of the object.}
\label{tabl:multiplets4}
\begin{tabular}{lr}
\hline
\hline
\multicolumn{1}{c}{TIC ID (other ID)} & \multicolumn{1}{c}{P$_{rot}$} \\
 &  \multicolumn{1}{c}{[h]} \\
\hline
0011626063 (GD~385) & 93.2 \\
0313109945 (GD~491) & 85.5 \\
1102242692 (WD~1526+558)& 25.1 \\
1103347368 (WD~1521-003) & 10.0 \\
\hline
\end{tabular}
\end{table}

Tables~\ref{tabl:multiplets3} and \ref{tabl:multiplets4} present the comparisons of our rotation periods with the previous results and the new detections, respectively. As Table~\ref{tabl:multiplets4} shows, we were able to determine the rotation periods for four stars of our sample for the first time.

We compare our results with literature data in Fig.~\ref{fig:rotmass}. Our results are in agreement with the former findings that larger-mass WDs rotate faster than their lower-mass counterparts. Notably, our two fastest rotators TIC~0007675859 and G~226-29 have the largest stellar masses.

For completeness, we also calculated the rotation rates for one star, assuming $3\,\ell=2$ rotationally split modes. We considered two cases: adjacent $m$ components or $m=-2,0,2$ components. The selected star for these calculations was TIC~0101014997 (BPM~31594), which shows a well-detected triplet with an average frequency separation of $12.415\,\mu$Hz. Considering Eq.~\ref{eq:rot}, \mbox{$C_{k,\ell} \approx 1/\ell(\ell+1)$}, and in the case of $\ell=2$ modes, \mbox{$C_{k,\ell} \eqsim 0.17$}. These values lead to a $18.6$\,h rotation period assuming that the triplet frequencies are adjacent $m$ components. Considering the possibility that the three frequencies are $m=-2,0,2$, the rotation period of TIC~0101014997 would be $37.3$\,h, which is two times that in the case of adjacent $m$ modes. For $\ell=1$ modes, we obtained a rotation period of  $11.9$\,h; but assuming $\ell=2$ modes, longer rotational periods are expected.

%%%%%%%%%%%%%%%%%%%%%%%%%%%%%%%%%%%%%%%%%%%%%%%%%%%%%%%%%%%%%%%%%%%%%%%%%%%%%%%%%%%%%%%%%%%%%%%%%%%%%%%%%%%

\begin{acknowledgements}

We thank the anonymous referee for the constructive comments and recommendations on the manuscript. We also thank the helpful remarks on our manuscript of both Keaton J. Bell (Department of Physics, Queens College, City University of New York) and J.~J. Hermes (Department of Astronomy \& Institute for Astrophysical Research, Boston University).

ZsB and \'AS acknowledge the financial support of the Lend\"ulet Program of the Hungarian Academy of Sciences, projects No. LP2018-7/2022. This research was supported by the KKP-137523 `SeismoLab' \'Elvonal grant of the Hungarian Research, Development and Innovation Office (NKFIH).

ZsB acknowledges the support by the J\'anos Bolyai Research Scholarship of the Hungarian Academy of Sciences.

This paper includes data collected with the TESS mission, obtained from the MAST data archive at the Space Telescope Science Institute (STScI). Funding for the TESS mission is provided by the NASA Explorer Program. STScI is operated by the Association of Universities for Research in Astronomy, Inc., under NASA contract NAS 5–26555.

\end{acknowledgements}

%-------------------------------------------------------------------

%\begin{thebibliography}{}
%\end{thebibliography}

\bibliographystyle{aa} % style aa.bst
\bibliography{rotation} % your references Yourfile.bib

\end{document}